# Dark state as a measurable state by a dispersive readout without a Purcell limit

Wei-Chen Chien[1, †], Jyh-Yang Wang[2, †], Yen-Yu Chiang[3], Cheng-Chengh Huang[3] ,Lih-Chieh Hsaio[3], Yen-Chun Chen[3], Cen-Shawn Wu[3,4], Chiidong Chen[3], and Watson Kuo[1,3,*]

*[1]Department of Physics, National Chung Hsing University, Taichung, Taiwan*

*[2]Department of Electrical Engineering, Feng Chia University, Taichung, Taiwan*

*[3]Research Center for Critical Issues, Academia Sinica, Tainan, Taiwan*

*[4]Department of Physics, National Changhua University of Education, Changhua, Taiwan*

It is believed that the enhancement in qubit-resonator coupling allows a better dispersive readout but introduces greater Purcell loss. In this work, we propose that a dark mode in a coupled quantum system may violate this rule by introducing the ZZ interaction between the dark and a bright mode. The dark mode may exhibit an effective zero coupling strength, and zero Purcell loss with the resonator photons. Nevertheless, the dispersive shift is almost the same as the bright state, due to the higher order perturbation introduced by the higher excited states. Such a state demonstrates the measurability without a Purcell limit in circuit quantum electrodynamics.

## Introduction

Suppressing environmental decoherence is a central challenge in quantum electrodynamics and a prerequisite for scalable quantum-computing architectures.[1,2] For initialization, control, and readout, quantum systems must remain coupled to external degrees of freedom, which simultaneously introducing additional dissipation channels[3]. This tension is especially evident in superconducting circuit quantum electrodynamics, where a qubit is dispersively coupled to a resonator so that its state produces a state-dependent shift of the resonator [4-6]. However, the exchange interaction underlying this dispersive coupling also makes the qubit radiatively bright to the resonator and therefore exposes it to Purcell decay[5]. This creates an inherent trade-off between readout distinguishability and qubit lifetime[7]. Although schemes have been made to engineer the readout device for suppressing the qubit loss, it only preserves the lifetime of qubits in a narrow frequency range[8].

[†] These authors contributed equally.

[*] Corresponding author. E-mail address: wkuo@phys.nchu.edu.tw

A natural route beyond this trade-off is to create qubit states that remain measurable while being linearly decoupled from the resonator. This possibility raises two closely related questions: how can destructive interference produce such a radiatively decoupled state, and how can the state remain observable despite its suppressed coupling to the readout channel? Inspired by the concept of a decoherence-free subspace[9], the first question is understood within the framework of subradiance and dark-state formation. The concept of subradiant states is rooted in Dicke's seminal work and subsequent studies of emitters coupled to a common electromagnetic environment [10-13]. Interference among emission amplitudes within shared radiative channels produces collective modes with enhanced or suppressed decay, with complete cancellation yielding a radiatively dark state. Such collective states have been explored from two-emitter systems to many-body atomic[14,15] and superconducting platforms [16-18]. Related dark-state mechanisms also arise from interference between internal transitions or spatially separated coupling points [19,20].

The second question is particularly important and has motivated architectures that separate radiative coupling from dispersive visibility. A three-island transmon is proposed for realizing this separation through bright dipole-like and dark quadrupole-like modes [21]. The dark mode retains an indirect dispersive response through its interaction with the bright mode. Subsequent experiments demonstrated tunable resonator coupling and coherent control in the weak-coupling regime[22,23]. More recently, the P-mon extended this strategy by introducing an auxiliary mediator mode that provides cross-Kerr-based readout while the protected mode remains linearly decoupled from the resonator[24].

Although these works established the feasibility of combining radiative protection with dispersive readout, the underlying mechanism is embedded in relatively complex multimode architectures. Here, we isolate this mechanism in a minimal and experimentally transparent circuit-QED setting with a simple toy model as shown in Fig. 1(a). The dark mode, despite its strongly suppressed exchange coupling to the resonator, remains visible through its ZZ interaction[25-27] with the bright mode. The model is described by the Hamiltonian

$$\frac{H}{\hbar} = \omega_r \hat{n}_r + \omega_b \hat{n}_b + \frac{\alpha_b}{2}\hat{n}_b(\hat{n}_b - 1) + \omega_d \hat{n}_d + \frac{\alpha_d}{2}\hat{n}_d(\hat{n}_d - 1) + g_{br}\left(\hat{a}_r^\dagger \hat{a}_b + \hat{a}_r \hat{a}_b^\dagger\right) - 2g_{zz}\hat{n}_b\hat{n}_d. \quad (1)$$

Here, $\hat{a}_r$, $\hat{a}_b$, and $\hat{a}_d$ are the lowering operators, $\hat{n}_r$, $\hat{n}_b$, and $\hat{n}_d$ are the number operators and $\omega_r$, $\omega_b$, and $\omega_d$ are mode frequencies of the resonator, bright mode, and dark mode, respectively. $g_{br}$ describes the exchange coupling between the photon and the bright mode, while $g_{zz}$ describes the $ZZ$-type interaction between the bright and dark modes. Denoting the bare product basis as $|n, j, k\rangle$, where $n$, $j$ and $k$ are the quantum numbers of the photon, the bright and dark modes, respectively, the non-interacting energies are

$$\frac{E_{n,j,k}}{\hbar} = n\omega_r + j\omega_b + \frac{\alpha_b}{2}j(j-1) + k\omega_d + \frac{\alpha_d}{2}k(k-1) - 2g_{zz}jk. \quad (3)$$

The photon-bright mode exchange coupling, $V = \hbar g_{br}(\hat{a}_r^\dagger \hat{a}_b + \hat{a}_r \hat{a}_b^\dagger)$, leaves the dark-mode index $k$ conserved. Therefore, the energy corrections can be analyzed independently within each state manifold defined by $n + j = \text{const.}$

As illustrated in Fig. 1(b), the bare state $|n, j, k\rangle$ is repelled by the states $|n-1, j+1, k\rangle$ and $|n+1, j-1, k\rangle$ within the same rotating-wave approximation (RWA) manifold. According to the second-order perturbation theory, the frequency shift is given by

$$\begin{aligned}\delta E_{n,j,k} &= \frac{|\langle n-1, j+1, k|V|n, j, k\rangle|^2}{E_{n,j,k} - E_{n-1,j+1,k}} + \frac{|\langle n+1, j-1, k|V|n, j, k\rangle|^2}{E_{n,j,k} - E_{n+1,j-1,k}} \\ &= -\hbar\frac{n(j+1)g_{br}^2}{\Delta_{br} + j\alpha_b - 2g_{zz}k} + \hbar\frac{j(n+1)g_{br}^2}{\Delta_{br} + (j-1)\alpha_b - 2g_{zz}k} \quad (4)\end{aligned}$$

with $\langle n-1, j+1, k|V|n, j, k\rangle = \hbar g_{br}\sqrt{n}\sqrt{j+1}$ and $\Delta_{br} \equiv \omega_b - \omega_r$. When the dark mode number is $k$, the resonator transition frequency conditioned on the bright and dark levels is $\omega_r^{(j,k)} = (\tilde{E}_{n+1,j,k} - \tilde{E}_{n,j,k})/\hbar$, acquiring a dark-state-dependent dispersive shift, even though the dark mode has no direct exchange coupling to the resonator.

The same mechanism can be viewed from the complementary perspective shown in Fig. 1(c). By comparing the corrected energies of the $k = 0$ and $k = 1$ at fixed $n$ and $j$, the photon-number-dependent dark-mode transition frequency is $\omega_d^{(n,j)} = (\tilde{E}_{n,j,1} - \tilde{E}_{n,j,0})/\hbar$, depending on photon number $n$. Thus, the same perturbative process can be interpreted in two equivalent ways: as a dark(bright)-state-dependent dispersive shift of the resonator, or as a resonator-photon-number-dependent AC Stark shift of the dark(bright) mode. By denoting bright-dark detuning $\Delta_{br} = \omega_b - \omega_r$, the previous results are $\chi_b = 2g_{br}^2\alpha_b/\Delta_{br}(\Delta_{br} + \alpha_b)$ and $\chi_d = -2g_{br}^2 g_{zz}/\Delta_{br}(\Delta_{br} - 2g_{zz})$, while the later results are $\delta\omega_b = 2n_r g_{br}^2\alpha_b/\Delta_{br}(\Delta_{br} + \alpha_b)$ and $\delta\omega_d = -2n_r g_{br}^2 g_{zz}/\Delta_{br}(\Delta_{br} - 2g_{zz})$.

We propose that a coupled transmon pair with spatial symmetry sharing a common readout resonator can realize this target Hamiltonian. Commonly referred to as a "dual-rail qubit", such a symmetric transmon architecture is particularly advantageous because single-photon loss events manifest predominantly as detectable erasure errors. [28-31]. In contrast to conventional designs where individual readout resonators are coupled to each transmon, our architecture employs a single readout resonator symmetrically coupled to both transmons. This configuration naturally supports

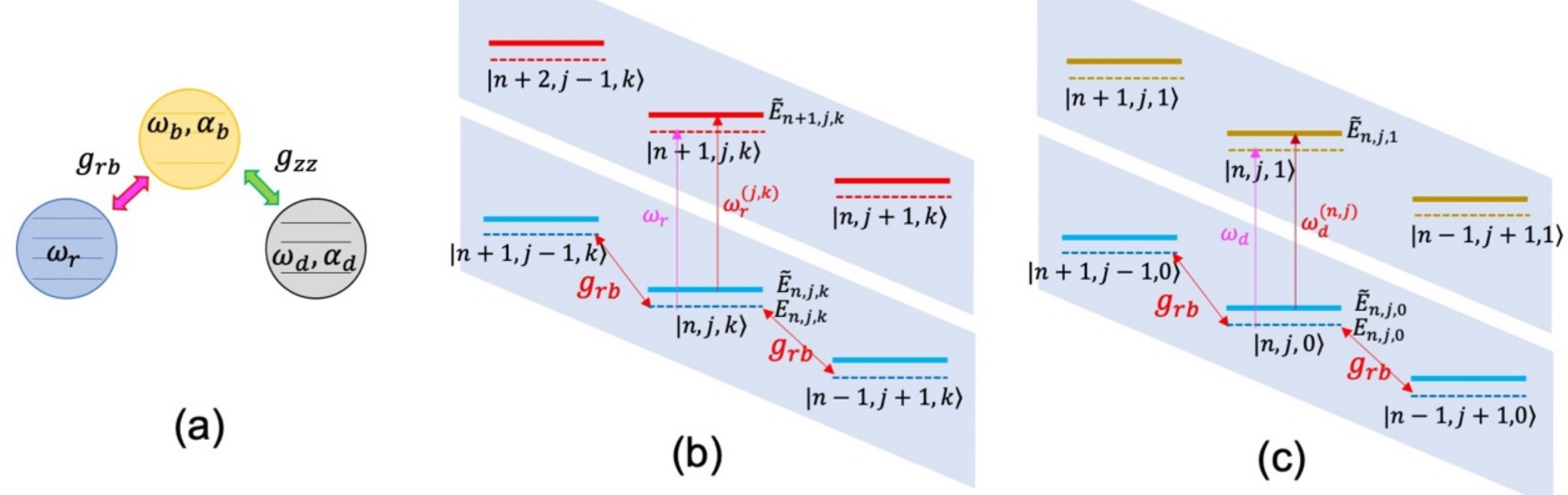

Fig. 1. (a) The proposed 3-body model. (b) The light-blue shaded regions denote neighboring RWA manifolds with fixed dark-mode level $k$, where the conserved excitation number is $n+j$. Solid lines represent bare product states $n, j, k\rangle$, while dashed lines indicate the second-order corrected energies $\tilde{E}_{n,j,k}$. The resonator transition frequency is $k$-dependent through the *ZZ*-shifted bright-resonator detuning. (c) The photon-number-dependent AC Stark shift of the dark mode.

orthogonal collective modes, namely an even ("bright") mode and an odd ("dark") mode.

With an appropriate transformation, the system can be recast into a set of coupled resonator($r$), bright-mode($b$), and dark-mode($d$) oscillators. We denote $\Phi_k$ as the generalized coordinate "position" and introduce its conjugate momentum $Q_k$ for $k = \{r, b, d\}$. Following the standard quantization procedure, the Hamiltonian with applying the RWA resembles that described by Eq. (1) by using the capacitance matrix involving Q1, Q2 and resonator electrodes [32,33]. With the $\{r, b, d\}$ basis, we obtained $g_{ZZ} = \sqrt{\alpha_b \alpha_d}$, in which $\alpha_b$ and $\alpha_d$ are the anharmonicity of $b$ and $d$ states. The detailed derivation can be found in supplemental document. An important observation is that $g_{ZZ}$ is on the order of the charging energy of the transmon and similar to the ordinary resonator-qubit coupling strength, $g_{br}$. For studying the Purcell limit, we employed a master equation calculation for the system initially prepared at the excited state $|b\rangle$ and $|d\rangle$ with a phenomenological relaxation rate on the resonator to represent the loss from the readout port.

Experimentally, we demonstrate that this antisymmetric collective mode combines an enhanced lifetime with a measurable resonator response. Our results provide a direct and conceptually transparent realization of a readable, Purcell-protected dark mode, and expose the physical origin of the separation between radiative coupling and dispersive visibility.

## Experimental Method

Fig. 2(a) shows a photograph of the full chip. The quantum system comprises individual X-mon qubits[34,35], each equipped with dedicated Z (flux bias) and X (microwave drive) lines to provide enhanced flexibility for flux tuning and coherent control. The artificial atoms are coupled to a $\lambda/4$ coplanar waveguide (CPW) resonator with a fundamental resonant frequency of 5.847(bare) 5.849(dressed) GHz, which is further coupled to a readout CPW transmission line for state measurement. As shown in the magnified view in Fig. 2(b), these coupled superconducting artificial atoms are flux-tunable X-mons fabricated from aluminum (Al). All devices were patterned by a single-step electron-beam lithography, Al metallization, followed by lift-off. Josephson junctions were made by Dolan technique during the two-angle Al depositions. To extract the capacitance matrix elements between the electrodes, a boundary-element method simulation was performed, the details of which are provided in the Supplementary Information. The coupling energy between the resonator and an individual X-mon is estimated to be $g/h = 53$ MHz, while the inter-Xmon coupling is $g_{12}/h = 211$ MHz. Key energy scales of the system are summarized in Table 1. The artificial atoms, denoted as Q1 and Q2, feature identical geometries, yielding nearly identical frequencies at zero magnetic field. The transmon frequencies $\omega_1$ and $\omega_2$ are detuned by 1 to 2 GHz from the readout resonator frequency, placing the system well within the dispersive readout regime. The sample was wire-bonded to a three-layer printed circuit board (PCB), packaged, and shielded using concentric copper and aluminum cylinders. The packaged device was cooled in a dilution refrigerator to a base temperature of 20 mK. Detailed schematics of the control lines, including the associated attenuators, filters, isolators, and amplifiers, are provided in the Supplementary Information. Pulsed microwave signals were generated and analyzed using a Keysight M3000 and a Quantum Machines OPX+ control platform.

Table 1 Important sample parameters. The units for $\omega_d$, $\omega_d$, and $\omega_r$ are GHz while others are MHz.

| $E_C$ | $\omega_d$ | $\omega_b$ | $\omega_r$ | $g_r$ | $g_{rb}$ | $g_{12}$ | $g_{zz}$ | $\alpha_b$ | $\alpha_d$ |
|---|---|---|---|---|---|---|---|---|---|
| 168 | 3.95 | 4.35 | 5.849 | 55.6 | 75.2 | 211 | 92.7 | -84.2 | -102.1 |

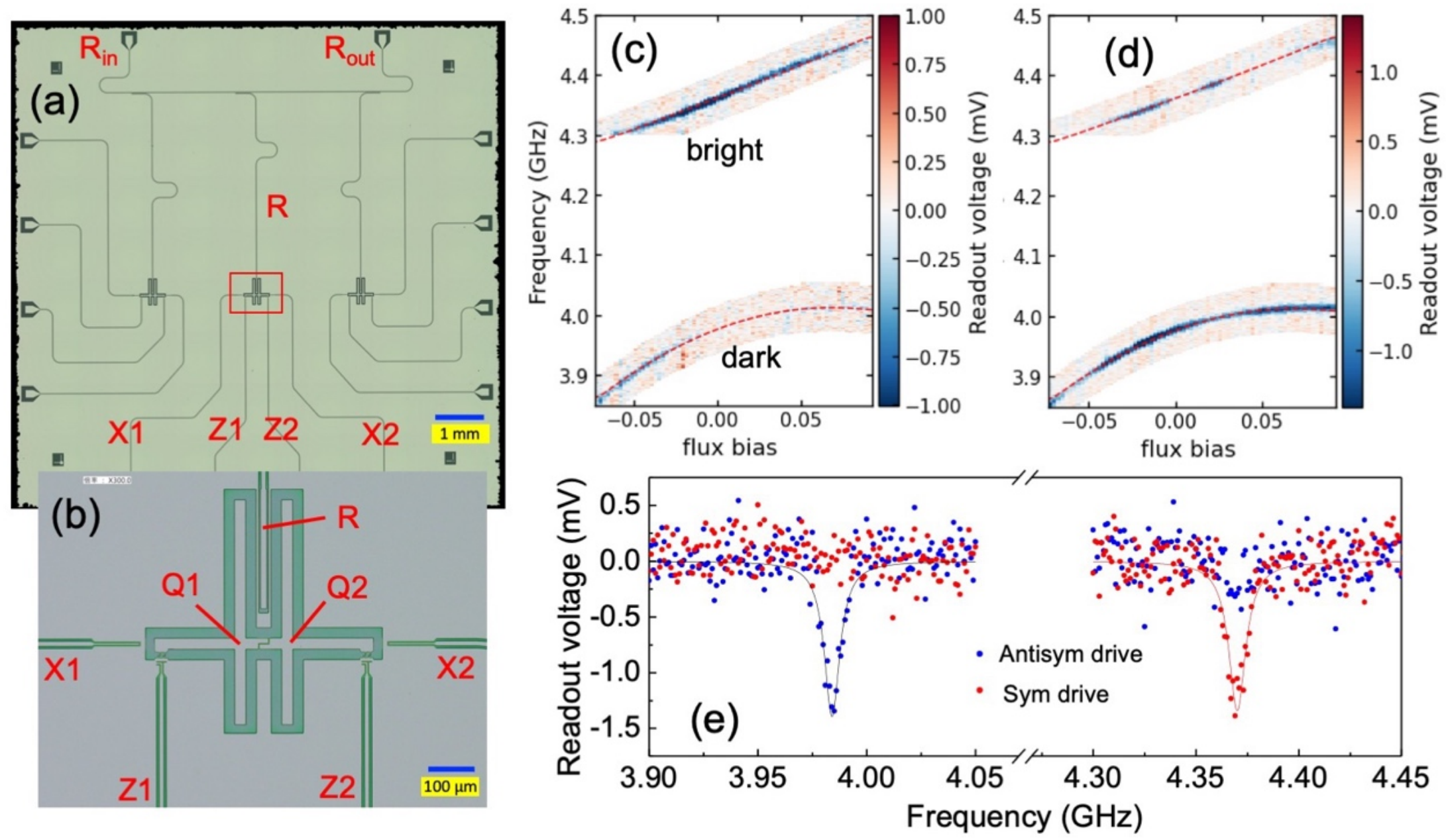

Fig. 1 (a) A photograph of two capacitively coupled X-mon qubits symmetrically coupled to a readout resonator(R). (b) Enlarged picture of the X-mons. (c)(d) Coupled transmons spectrum as a function of flux bias applied to Q1, with $\omega_2$ is 4.2 GHz. The zero-flux bias point denotes the Q1-Q2 degeneracy, where symmetric and antisymmetric modes are formed. The spectrum is obtained by symmetric (in-phase) driving (c) and antisymmetric (out-of-phase) driving (d). Symmetric driving selectively excites the bright state, while anti-symmetric one allows dark state transition. (e) The spectrum at degenerate point. The anti-symmetric mode at lower frequency cannot be excited with symmetric driving(red), while the symmetric mode at lower frequency cannot be excited with anti-symmetric driving(blue).

**Results and Discussions**

*Two-Tone Spectroscopy* Qubit spectroscopy was performed at a baseline frequency of $\omega_0 \sim 4.18$ while varying the flux bias of qubit 1. The resulting spectra for symmetric and antisymmetric microwave driving, achieved via the X1 and X2 drive lines, are shown in Figs. 2(c) and 2(d), respectively. Here zero flux is defined to be the flux change from the degenerate point, $\omega_1 = \omega_2$. Detailed calibration procedures for both driving configurations are provided in the Supplementary Information. Under symmetric driving, the resonance of the higher-frequency mode is clearly observable across a detuning range of $\omega_1 - \omega_2 = \pm 0.35$ GHz(flux bias change from -0.07 to 0.09), whereas absorption by the lower-frequency mode is negligible. Conversely,

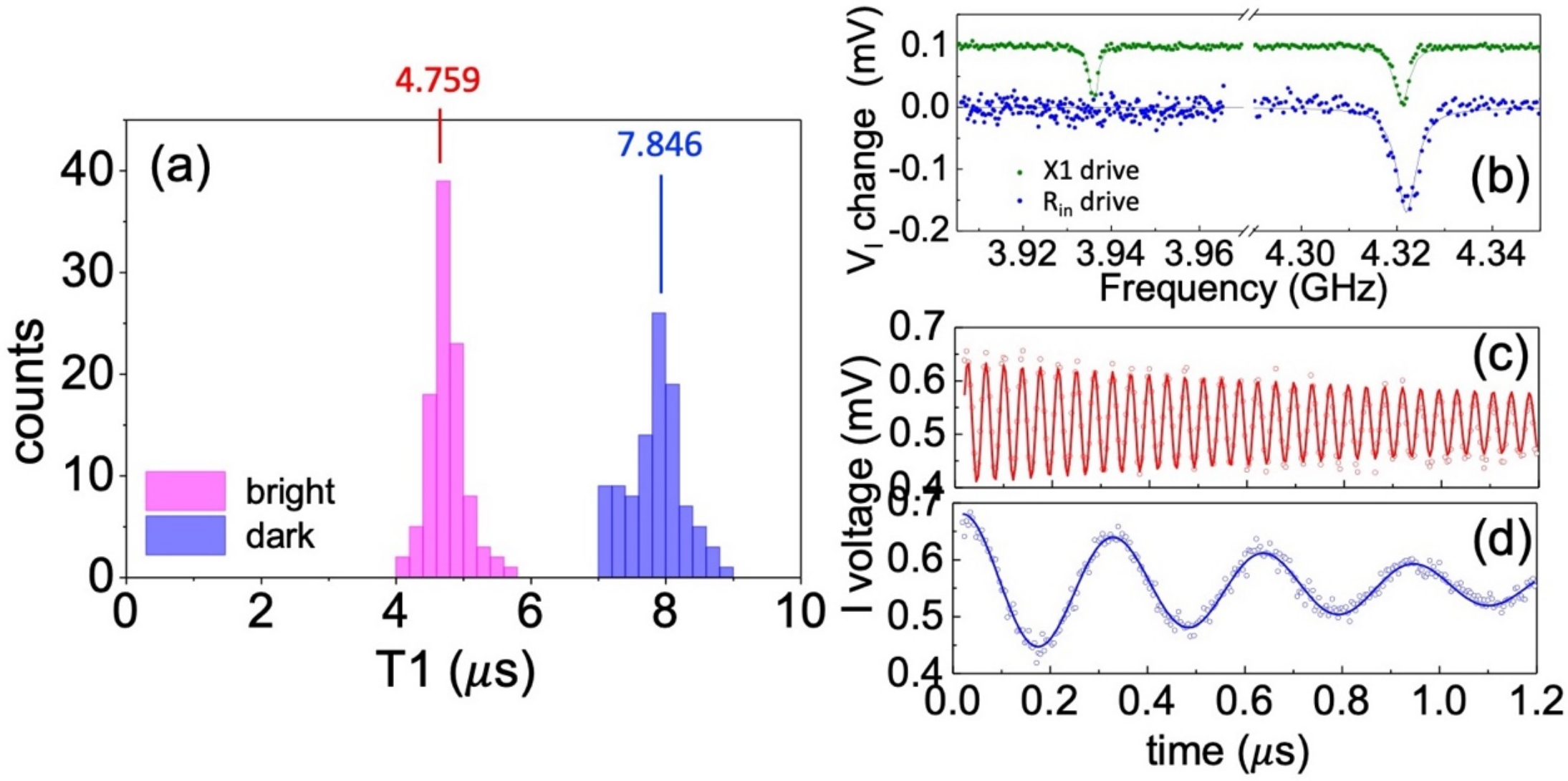

Fig. 3 (a) Histograms of T1 for bright(red) and dark states(blue) (b) Spectrum of the bright and dark states with a long driving tone applied from X1 gate(green) and $R_{in}$ line(blue). The green curve is shifted vertically for clarity. (c) The time Rabi oscillations of bright state driving from $R_{in}$ line. (d) The time Rabi oscillations of dark state driving from $R_{in}$ line. The microwave amplitude is 4× as that in (c).

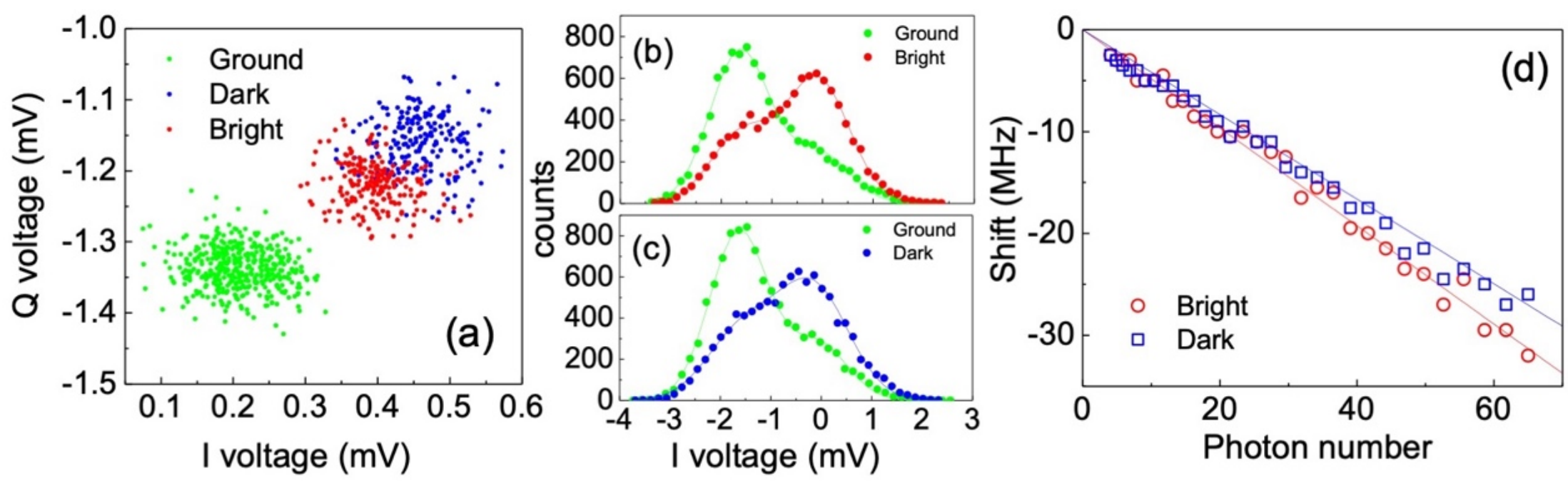

Fig. 4 (a) The IQ plots for 100× averaged readout for the states prepared at ground(green), bright(red) and dark(blue). (b)(c) The distribution of readout voltage of single-shot measurement when the state is prepared with $|g\rangle$, $|b\rangle$ and $|d\rangle$. (d) ac-Stark shifts for states $|d\rangle$(red) and $|b\rangle$(blue). Solid curves show the theoretical predictions.

antisymmetric driving reveals transitions to the lower-energy level while leaving the higher-energy level unexcited. These results experimentally confirm the symmetry inherent in these two-qubit states; specifically, when both transmons are driven by a common photon field, the higher and lower energy levels correspond to the bright state $|b\rangle$ and dark state $|d\rangle$, respectively. Fig. 2(e) further illustrates the suppression of dark and bright resonances under weak symmetric and antisymmetric driving, respectively.

*Relaxation times* Due to the symmetric coupling of the transmons to the readout resonator, the bright $|b\rangle$ and dark $|d\rangle$ states exhibit distinct photon interaction characteristics. Under conditions where qubit dissipation is dominated by the readout resonator and its peripheral circuitry, these distinct interactions are expected to manifest as observable differences in the relaxation times of $|b\rangle$ and $|d\rangle$. To investigate this effect, standard T1 measurements were performed using symmetric or antisymmetric XY drive configurations to ensure precise $\pi$-pulse excitations for each respective state. The measured relaxation times highlight the lower loss of the antisymmetric state, yielding average lifetimes of T1 =7.846 $\mu$s for $|d\rangle$ and T1 = 4.579 $\mu$s for $|b\rangle$. The measured T1 lifetimes differ by a factor of approximately 1.7, deviating significantly from the theoretical Purcell limit. This discrepancy is attributed to alternative energy dissipation mechanisms, such as materials-related dielectric loss, quasiparticle poisoning, and loss from XY lines. Nevertheless, microwave driving applied directly via the readout resonator can be utilized to probe the coupling between the resonator photons and the collective qubit states.[8] First, a 10 μs -long weak microwave drive was applied through the $R_{in}$ line to map the state spectrum. As shown in Fig. 3(b), a transmission change in the readout tone is exclusively observed at the bright-state resonance, whereas the dark state remains unaltered (blue symbols). For comparison, an identical measurement performed by applying the drive tone through the X1 line resolves both the bright and dark state resonances (green symbols). This feature confirms the suppressed interaction between the readout resonator and the dark state. Second, time-resolved Rabi oscillations were driven via the resonant microwave field from the $R_{in}$ line. As shown in Fig. 3(c), the Rabi frequency of the bright state is 26.9 MHz, whereas the dark state exhibits a Rabi frequency of 3.24 MHz when driven at a fourfold (4×) higher microwave amplitude. Scaled to a uniform driving amplitude, the ratio between the Rabi frequencies is 33.2, indicating that the Purcell lifetime limit of the dark state is approximately 33 times greater than that of the bright state.
*Dispersive shift and Ac-Stark shift* Notably, despite experiencing negligible relaxation via the resonator due to an effectively null photon interaction, the dark state can still be probed via dispersive readout, as demonstrated by our spectroscopy, Rabi, and T1 measurements. To confirm this dispersive shift, we performed single-shot measurements with the system prepared in the ground state $|g\rangle$, $|b\rangle$ and $|d\rangle$. Due to the low signal-to-noise ratio inherent to our readout resonator design, we averaged the readout IQ voltages 100 times for the data shown in Fig. 4(a). Both the dark (blue symbols) and bright (red symbols) states exhibit a clear separation from the ground state, whereas the dark and bright IQ distribution clouds closely overlap. Figs. 4(b) and 4(c) show the single-shot readout voltage distributions for the three prepared states. In all cases, the readout voltage distributions generally exhibit a double-Gaussian profile with distinct separation. The comparable voltage separation observed for both $|b\rangle$ and $|d\rangle$ indicates that the two states possess similar dispersive shifts. Because the direct dispersive shift is too small to be resolved with high precision, we investigate the AC Stark shifts of $|b\rangle$ and $|d\rangle$, which can be amplified by the intra-cavity photon number $n$ [36,37]. Experimentally, the photon number is controlled by the microwave drive power when applying an on-resonance pump tone to the resonator during two-tone spectroscopy. As illustrated in Fig. 4(d), both the $|b\rangle$ and $|d\rangle$ states exhibit highly similar AC Stark shifts. The minor discrepancy observed experimentally is attributed to a small difference in detuning $\Delta$; specifically, $|b\rangle$ has a smaller detuning $\Delta$ than $|d\rangle$, resulting in a slightly larger shift. The experimental data highly agree with the theoretical predictions -0.4818 and -0.4159 MHz per photon for $|b\rangle$ and $|d\rangle$, respectively.

**Conclusion**

In summary, we propose a general route to build a measurable state without linear radiative loss in the framework of circuit quantum electrodynamics. By introducing the ZZ type interaction between a dark and bright state, the dark state becomes visible by the readout resonator. The required ingredients are realized with strongly coupled transmons with a symmetric coupling strength to a readout resonator. We identify the bright and dark states by symmetric and anti-symmetric driving the system, and may unambiguous excite the system to each state. The relaxation time measurement, together with the Rabi oscillations driven from the resonator readout line, reveal that the dark state enjoyed a protection from Purcell loss due to the effective null resonator-mode interaction. With negligible effective coupling, the dark state population does introduce a finite frequency shift on the readout resonator through the ZZ interaction between the dark and bright states. AC-Stark shift better demonstrates the similar mode level shift with the distinct effective coupling strengths. These findings assure that the designed dark state enjoys the benefit of a measurable quantum state by the dispersive readout without a Purcell limit. The coupled transmon, also “dual-rail” architecture is attractive to be implemented with the scheme of erasure error detection. Our findings may find potential application in quantum state readout, information processing, and quantum state corrections.

**Funding.** National Science and Technology Council (114-2222-E-035-007, 115-2119-M-001-001, 115-2112-M-005-006)

**Acknowledgements.**

We are grateful for the computing support from the National Center for High-performance Computing and measurement support from the Instrument Center, NCHU.

We have used the facilities in Research Center of Critical Issues, Academia Sinica. Fruitful discussions with Y. F. Chen and C. T. Ke are acknowledged.

**Disclosures.** The authors declare that there are no conflicts of interest related to this article.

**Data availability.** Data underlying the results presented in this paper are not publicly available at this time but may be obtained from the authors upon reasonable request.